\documentclass[10pt,aps,prd,twocolumn,notitlepage,nofootinbib,floatfix,superscriptaddress,preprintnumbers]{revtex4-1}

\usepackage{amsmath, amsthm, amssymb, amsfonts, amsbsy, mathrsfs}

\usepackage{xcolor}
\usepackage{calligra,bm}
\usepackage{stmaryrd}

\usepackage{graphicx}
\graphicspath{{./images/}}

\usepackage[utf8]{inputenc}
\usepackage[export]{adjustbox}
\usepackage{wrapfig}

\usepackage[hidelinks,bookmarks=true]{hyperref} 
\hypersetup{pdfstartview=FitH,pdfhighlight=/O,colorlinks=false}

\begin{document}

\title{Minimal exponential measure model in the post-Newtonian limit}

\author{Justin C. Feng}
\affiliation{CENTRA, Departamento de F{\'i}sica, Instituto Superior T{\'e}cnico (IST), \\ Universidade de Lisboa (UL), Avenida Rovisco Pais 1, 1049 Lisboa, Portugal}

\author{Shinji Mukohyama}
\affiliation{Center for Gravitational Physics, Yukawa Institute for Theoretical Physics,\\ Kyoto University, Kyoto 606-8502, Japan}
\affiliation{Kavli Institute for the Physics and Mathematics of the Universe (WPI), \\The University of Tokyo Institutes for Advanced Study,\\The University of Tokyo, Kashiwa, Chiba 277-8583, Japan}

\author{Sante Carloni}
\affiliation{DIME Sezione Metodi e Modelli Matematici, Universit\`{a} di Genova,\\ Via All'Opera Pia 15, 16145 Genoa, Italy.}

\preprint{YITP-20-152, IPMU20-0122}


%
%

%
%
\begin{abstract} 
We examine the post-Newtonian limit of the minimal exponential measure (MEMe) model presented in [J. C. Feng, S. Carloni, Phys. Rev. D 101, 064002 (2020)] using an extension of the parameterized post-Newtonian (PPN) formalism which is also suitable for other type-I minimally modified Gravity theories. The new PPN expansion is then used to calculate the monopole term of the post-Newtonian gravitational potential and to perform an analysis of circular orbits within spherically symmetric matter distributions. The latter shows that the behavior does not differ significantly from that of general relativity for realistic values of the MEMe model parameter $q$. Instead the former shows that one can use precision measurements of Newton's constant $G$ to improve the constraint on $q$ by up to $10$ orders of magnitude.
\end{abstract}


%
%

\maketitle

%
%

%
%

\section{Introduction}
A recent article \cite{FengCarloni2019} introduced a class of generalized coupling theories (GCTs), the simplest of which was termed the minimal exponential measure (MEMe) model. These are modified theories of gravity that do not introduce new dynamical degrees of freedom; rather, they modify the interaction between spacetime and matter in a manner that preserves the Einstein equivalence principle (all matter is minimally coupled to an effective spacetime geometry). According to the classification scheme of \cite{Aoki2018,DeFelice2020,Aoki2020}, GCTs and the MEMe model are Type I minimally modified gravity (MMG) theories, since they only have two dynamical degrees of freedom and admit an Einstein frame (in the sense that the theories may be rewritten as general relativity [GR] with a modified source). While it was shown in \cite{FengCarloni2019} that the dynamical behavior of the MEMe model differs significantly from GR under the conditions present in the early Universe and within a matter distribution, the MEMe model reduces to GR in a vacuum---in this respect, the MEMe model is qualitatively similar to the Eddington-inspired Born Infeld (EiBI) theory \cite{BanadosFerreira2010}. However, the predictions of the MEMe model differ from those of GR within a matter distribution and in its coupling to matter. The purpose of the present article is to determine the degree to which these differences can be measured in the post-Newtonian limit.

Modified gravity theories, i.e. those that attempt to go beyond GR, have been extensively studied for at least three motivations: (i) to understand or solve mysteries in cosmology such as the origins of dark energy, dark matter, and inflation; (ii) to help develop the theory of quantum gravity; and (iii) to understand GR itself. Regarding (iii), even if GR is the genuine description of gravity in our Universe for a certain range of scales, the only way to prove it experimentally or observationally is to constrain possible deviations from GR by experiments or observations. In this regard, it is useful to have a universal parameterization of possible deviations from GR. For Solar System scales, the so-called parameterized post-Newtonian (PPN) formalism proved to be particularly useful. The standard PPN formalism includes ten parameters to parameterize deviations from GR and covers a wide range of gravitational theories beyond GR~\cite{Will2018}. However, there is no guarantee that the standard PPN formalism can be applied to all modified gravity theories. For example, in gravitational theories without the full diffeomorphism invariance, one cannot, in general, adopt the standard PPN gauge and thus may have to introduce additional PPN potentials or parameters (see e.g. \cite{Lin:2013tua}).  The MEMe model we consider here also requires an extension of the standard PPN formalism for a different reason: the nontrivial matter coupling inevitably generates potentials that are not included in the standard PPN formalism. These potentials are not only relevant in themselves, but they are also necessary to compute the standard ten PPN parameters.\footnote{From the bottom-up point of view, the ten PPN parameters are in principle independent. (Note, however, that there is one relation that is expected to hold for all reasonable theories~\cite{Will:1976zza}.) On the other hand, from the top-down point of view, once a specific theory of modified gravity is fixed, then the PPN parameters are written in terms of the parameters of the theory.}

In this article, we shall construct an extension of the PPN formalism appropriate for a subclass of type-I MMG and GCTs based on additional PPN potentials. We will then focus on the MEMe model, finding that one must add to the PPN metric a single new potential, which we denote $\Psi$, and some additional counterterms.
All of the counterterms are proportional to the pressure, mass density, and energy density for a fluid, so they vanish outside matter sources. This is, however, not the case for $\Psi$. Hence the PPN parameters for the MEMe model in the case of a test particle in an external field agree with those of GR except for the coefficient associated with $\Psi$.

In the context of the MEMe model, we find that the effects of the potential $\Psi$ can be absorbed into the Newtonian potential outside a matter distribution. This result suggests that the modification to the matter couplings can in the post-Newtonian limit be reinterpreted as a density-dependent modification of the gravitational constant $G$. Comparing with \cite{Lietal2018}, we argue that current laboratory methods can improve the constraint on the (single) parameter $q$ in the MEMe model by 10 orders of magnitude over the speed of light constraint discussed in \cite{FengCarloni2019}.

We also study circular orbits in the presence of spherically symmetric matter distributions and compare the predictions of the MEMe model with GR. Our findings suggest that in most astrophysical systems, the presence of a dilute matter distribution does not significantly affect the motion of matter in the MEMe model.

This paper is organized as follows. First, we review the MEMe model and GCTs in Sec.~\ref{sec:MEMe}. We then discuss the Newtonian limit in Sec.~\ref{sec:Newtonian} and develop the PPN formalism for the MEMe model in Sec.~\ref{sec:PPN}. Afterward, in Sec.~\ref{sec:monopole} we consider to post-Newtonian order the monopole term for the MEMe model and discuss how constraints on the variation of the effective gravitational constant may be used to constrain the parameter $q$ in the MEMe model. Finally, in Sec.~\ref{sec:counterterms} we compare the behavior of circular orbits within a spherically symmetric matter distribution in the MEMe model to that of GR. Section~\ref{sec:summary} is then devoted to a summary of the paper and some discussions.


%
%

%
%

\section{Generalized coupling theories and the MEMe model}
\label{sec:MEMe}

Generalized coupling theories are defined by an action of the form \cite{FengCarloni2019}
\begin{equation}\label{GCA-GENAction}
\begin{aligned}
S_{GC}= \int d^4x \biggl[& \frac{1}{2 \kappa}\left(R - 2 \, [\Lambda - \lambda(1-F)]\right)\sqrt{-{g}} \\
& + L_{m}[\phi,\mathfrak{g}^{\cdot\cdot}] \sqrt{-\mathfrak{g}} \biggr],
\end{aligned}
\end{equation}
\noindent where the metric $\mathfrak{g}_{\mu \nu}$ is assumed to have the form
\begin{equation}\label{GCA-AuxiliarymetricGC}
\mathfrak{g}_{\mu \nu} = \Xi(A{_\cdot}{^\cdot}) \,  A{_\mu}{^\alpha} \, A{_\nu}{^\beta} \, g_{\alpha \beta},
\end{equation}

\noindent and the function $F=F(A{_\cdot}{^\cdot})$ is chosen so that in a vacuum $A{_\mu}{^{\alpha}}=\delta{_\mu}{^{\alpha}}$ is an extremum of the action. Upon varying the action with respect to the metric and remembering that $A{_\mu}{^\alpha}$ is independent of $g_{\mu\nu}$, one obtains field equations of the form
\begin{equation}\label{GCA-GEN-GFE}
 G_{\mu \nu} +\left[  \Lambda- \, \lambda\left(1 - F\right)\right] \, g_{\mu \nu} = \kappa \, \Xi \, |A{_\cdot}{^\cdot}| \, \bar{A}{^\alpha}{_\mu} \, \bar{A}{^\beta}{_\nu} \, \mathfrak{T}_{\alpha \beta},
\end{equation}
\begin{equation}\label{GCA-GEN-AFE}
(\delta{_\mu}{^{\alpha}}-A{_\mu}{^{\alpha}})f^{\nu}_{\alpha}={\Xi^2 \, |A{_\cdot}{^\cdot}|}  \, \left[\mathfrak{T}_{\alpha \beta} \, \mathfrak{g}^{\mu (\alpha} \, \bar{A}{^{\beta)}}{_\nu} + \mathfrak{T} \, \frac{1}{2 \, \Xi} \, \frac{\partial \Xi}{\partial A{_\mu}{^\nu}} \right],
\end{equation}

\noindent where $\bar{A}{^{\alpha}}{_\mu}$ is the inverse of $A{_\mu}{^\alpha}$ and $f^{\nu}_{\alpha}=f^{\nu}_{\alpha}(A{_\cdot}{^\cdot})$.

The MEMe model, discussed at length in \cite{FengCarloni2019}, is a simple example of a generalized coupling theory. The MEMe model is defined by the following action:
\begin{equation}\label{GCA-MEMeAction}
\begin{aligned}
S[\phi,g^{\cdot\cdot},A{_{\cdot}}{^{\cdot}}]= \int d^4x \biggl\{& \frac{1}{2 \kappa}\left[R - 2 \, \tilde{\Lambda} \right]\sqrt{-{g}} \\
&+ \left(L_{m}[\phi,\mathfrak{g}^{\cdot\cdot}] - \frac{\lambda}{\kappa} \right) \sqrt{-\mathfrak{g}} \biggr\} ,
\end{aligned}
\end{equation}

\noindent where $\kappa := 8 \pi G$ and the Jordan-frame metric $\mathfrak{g}_{\mu \nu}$ is defined (with $A:=A{_\sigma}{^\sigma}$) as
\begin{equation}\label{GCA-JordanMetric}
\mathfrak{g}_{\mu \nu} = e^{(4-A)/2} \,  A{_\mu}{^\alpha} \, A{_\nu}{^\beta} \, g_{\alpha \beta} ,
\end{equation}

\noindent and $\tilde{\Lambda}=\Lambda - \lambda$, with $\Lambda$ being the observed value of the cosmological constant. Unless stated otherwise, indices are raised and lowered using the metric $g^{\mu \nu}$ and $g_{\mu \nu}$. Defining the parameter
\begin{equation}\label{GCA-qparameter}
q:=\frac{\kappa}{\lambda},
\end{equation}

\noindent the equation of ``motion'' for $A{_\mu}{^\alpha}$ takes the following form:
\begin{equation}\label{GCA-ExpFEs}
\begin{aligned}
{A}{_\beta}{^\alpha} - \delta{_\beta}{^\alpha} = q \left[ (1/4) \mathfrak{T} \, {A}{_\beta}{^\alpha} - \mathfrak{T}_{\beta \nu} \, \mathfrak{g}^{\alpha \nu} \right],
\end{aligned}
\end{equation}

\noindent where $\mathfrak{T}_{\mu \nu}$ is the energy-momentum tensor defined by the functional derivative of $\int L_{m}[\phi,\mathfrak{g}^{\cdot\cdot}] \sqrt{-\mathfrak{g}} \, d^4x$, and $\mathfrak{T}:=\mathfrak{g}^{\mu \nu}\mathfrak{T}_{\mu \nu}$. Here, we assume $q \mathfrak{T} \neq 4$. Since Eq. \eqref{GCA-ExpFEs} is an algebraic equation for $A{_\mu}{^\alpha}$, the tensor $A{_\mu}{^\alpha}$ does not introduce additional dynamical degrees of freedom. The trace of Eq. \eqref{GCA-ExpFEs} implies $A=A{_\sigma}{^\sigma}=4$. The gravitational equations are (setting $A=4$)
\begin{equation}\label{GCA-GEN-GFE-MEMe}
 G_{\mu \nu} +\left[  \Lambda- \, \lambda\left(1 - |A{_\cdot}{^\cdot}| \right)\right] \, g_{\mu \nu} = \kappa \, |A{_\cdot}{^\cdot}| \, \bar{A}{^\alpha}{_\mu} \, \bar{A}{^\beta}{_\nu} \, \mathfrak{T}_{\alpha \beta},
\end{equation}

\noindent where $\bar{A}{^\alpha}{_\mu}$ is the inverse of $A{_\mu}{^\alpha}$ as already explained and $|A{_\cdot}{^\cdot}|=\det(A{_\cdot}{^\cdot})$. One may see from the form of Eq. \eqref{GCA-GEN-GFE-MEMe} that the MEMe model admits an Einstein frame in the sense of \cite{Aoki2018}, making this a type-I MMG. Here, the operating definition for an \textit{Einstein frame} is a choice of variables in which a theory is recast as GR with a modified source, which may involve additional degrees of freedom. We define the \textit{Jordan frame} as a choice of variables in which matter is minimally coupled to the metric tensor. In the MEMe model, it is the frame in which matter is coupled to the metric tensor $\mathfrak{g}_{\mu \nu}$. We should stress however that, despite some similarities, these frames are not related to the well-known conformal transformations in modified gravity. The choice of frame is important also because it specifies the worldlines of free-falling test particles: since matter is minimally coupled to the Jordan-frame metric, one expects that small clumps of matter follow the worldlines of test particles as defined by the Jordan-frame metric.\footnote{One should keep in mind that since ${A}{_\beta}{^\alpha} = \delta{_\beta}{^\alpha}$ in a vacuum, the Einstein- and Jordan-frame metrics coincide in the absence of matter.} For this reason, the Jordan frame is the most physically relevant choice.

Equation \eqref{GCA-ExpFEs} can be solved exactly for a single perfect fluid. The dual (lowered-index) fluid four-velocity $u_\mu$ is constructed from the gradients of the potentials, so it is appropriate to regard $u_\mu$ to be the metric-independent fluid variables. The energy-momentum tensor for the fluid takes the form
\begin{equation} \label{GCA-EnergyMomentumPerfectFluid}
\mathfrak{T}_{\mu \nu} = \left(\underline{\rho} + p\right)u_\mu u_\nu + p \> \mathfrak{g}_{\mu \nu},
\end{equation}

\noindent the Jordan-frame trace of which is $\mathfrak{T}=3 p-\underline{\rho}$. Note that, while $\mathfrak{g}^{\mu \nu} u_\mu u_\nu = -1$, $g^{\mu \nu} u_\mu u_\nu \neq  -1$. It is useful also to define a dual four-velocity vector which is normalized with respect to the Einstein-frame metric $g_{\mu \nu}$. Defining $\varepsilon := g^{\mu \nu} u_\mu u_\nu$, one can obtain such a four-velocity (defining $u^\mu := g^{\mu \nu} u_\nu$):
\begin{equation}\label{GCA-UnitFlowField}
U^\mu := {u^\mu}/{\sqrt{-\varepsilon}} ,
\end{equation}

\noindent where $u^{\mu}=g^{\mu\nu}u_{\nu}$, and it follows that $u_\mu \, u_\nu = - \varepsilon U_\mu \, U_\nu$.

Since the MEMe model admits two metric tensors $g_{\mu \nu}$ and $\mathfrak{g}_{\mu \nu}$, one should be careful when raising and lowering the indices of the four-velocity---while the (dual) vector $u_\mu$ is the lowered index Jordan-frame four-velocity, the raised index Jordan-frame four-velocity $\mathfrak{u}^\mu$ is defined as the following:
\begin{equation}\label{GCA-FourVelocityDefinition}
\mathfrak{u}^\mu := \mathfrak{g}^{\mu \nu} \, u_\nu ,
\end{equation}

\noindent which is in general not equal to ${u}^\mu$. One may obtain a simple relationship between the respective raised and lowered components of the Jordan-frame fluid four-velocity $\mathfrak{u}^\mu$ and $u_\nu$ by first noting that $A{_\mu}{^\alpha} u_\alpha \propto u_\mu$ and $\bar{A}{^\alpha}{_\mu} u_\alpha \propto u_\mu$; it follows that $\mathfrak{u}_\mu \propto u_\mu$ (where $\mathfrak{u}_\mu = g_{\mu \nu} \mathfrak{u}^\nu$). One may then write $\mathfrak{u}_\mu = a \, u_\mu$ where $a$ is some factor. Now recall that $u_\mu u^\mu = \varepsilon$, and since $u_\mu \mathfrak{u}^\mu = u_\mu \, u_\mu \, \mathfrak{g}^{\mu \nu} = -1$, one can show that $a = - 1/\varepsilon$ and obtain the result
\begin{equation} \label{GCA-JordanFrame4v}
\begin{aligned}
u_\mu = - \varepsilon \, \mathfrak{u}_\mu,
\end{aligned}
\end{equation}
\begin{equation} \label{GCA-EinsteinFrame4v}
\begin{aligned}
U^\mu = \sqrt{- \varepsilon} \, \mathfrak{u}^\mu.
\end{aligned}
\end{equation}

\noindent It follows that $\mathfrak{u}_\mu \mathfrak{u}^\mu = 1/\varepsilon$, and $U_\mu \, U_\nu = - \varepsilon \, \mathfrak{u}_\mu \, \mathfrak{u}_\nu$.

Given the following ansatz for $A{_\mu}{^\alpha}$
\begin{equation}\label{GCA-AnsatzRS}
A{_\mu}{^\alpha} = {Y} \, \delta{_\mu}{^\alpha} - \varepsilon \, Z \, U{_\mu} \, U{^\alpha},
\end{equation}

\noindent one can easily solve Eq. \eqref{GCA-ExpFEs}, with the result:
\begin{equation}\label{GCA-ASoln}
\begin{aligned}
Y &= \frac{4 (1 - p \, q)}{4 - q \,  (3 \, p - \underline{\rho})} \\
Z &= - \frac{q \, (p + \underline{\rho}) [4 - q \, (3 \, p - \underline{\rho})]}{4 \, (q \, \underline{\rho} + 1)^2}\\
\varepsilon &= - \frac{16 \, (q \, \underline{\rho} + 1)^2}{[4 - q \, (3 \, p - \underline{\rho})]^2}.
\end{aligned}
\end{equation}

\noindent The inverse $\bar{A}{^\alpha}{_\mu}=[\delta{_\mu}{^\alpha}+\varepsilon Z(Y+\varepsilon Z)^{-1}U{_\mu}U{^\alpha}]/Y$ has a similar form. The gravitational equation \eqref{GCA-GEN-GFE-MEMe} takes the form
\begin{equation}\label{GCA-GEN-GFE-EF}
G_{\mu \nu}=\kappa \, T_{\mu \nu},
\end{equation}
where $T_{\mu \nu}$ is the effective energy-momentum tensor in the Einstein frame defined by
\begin{equation}\label{GCA-ExpTmnEffDecomp}
T_{\mu \nu} = \left(\tau_1 + \tau_2\right) \, U_\mu \, U_\nu + \tau_2 \, g_{\mu \nu} ,
\end{equation}

\noindent and
\begin{equation}\label{GCA-ExpTmnEffDecompTs}
\begin{aligned}
\tau_1 & = |A{_\cdot}{^\cdot}| \, (p + \underline{\rho}) - \tau_2 , \\
\tau_2 & = \frac{|A{_\cdot}{^\cdot}| \, (p \, q - 1) + 1}{q}-\frac{\Lambda}{\kappa},
\end{aligned}
\end{equation}

\noindent with the following expression for the determinant:
\begin{equation}\label{GCA-ExpAdet}
|A{_\cdot}{^\cdot}| = \det(A{_\cdot}{^\cdot})=\frac{256 \, (1 - p \, q)^3 (q \, \underline{\rho} + 1)}{[4 - q \, (3 p - \underline{\rho}) ]^4}.
\end{equation}

So far, the gravitational field equations \eqref{GCA-GEN-GFE}, \eqref{GCA-GEN-GFE-MEMe}, and \eqref{GCA-GEN-GFE-EF} are written as dynamical equations for the metric tensor $g_{\mu \nu}$. One can in principle attempt to reexpress the field equations in terms of the metric $\mathfrak{g}_{\mu \nu}$. This can be done by solving Eq. \eqref{GCA-JordanMetric} for ${g}_{\mu \nu}$ and inserting the resulting expression into the Einstein tensor to obtain an expression for the gravitational field equations in the Jordan frame. In this case the resulting field equation will contain derivatives up the second order of the tensor $A{_\mu}{^\alpha}$. We do not report here the form of such an equation which is rather long. However, we wish to highlight this feature of the Jordan-frame field equations as it is relevant for the following discussion on the distinction between the MEMe model and other modified gravity theories and also the extension of the PPN formalism that we will present in the next section.

It is perhaps appropriate to summarize here some properties and features of the MEMe model. The tensor $A{_\mu}{^\alpha}$ is an auxiliary field satisfying an algebraic equation \eqref{GCA-ExpFEs}, so it does not contain additional dynamical degrees of freedom. The MEMe model therefore does not introduce dynamical instabilities beyond those already present in general relativity (such as Jeans instability). However, if one imagines the coupling tensor $A{_\mu}{^\alpha}$ to be a coarse-grained description for dynamical degrees of freedom, then one can treat the term containing $\lambda=\kappa/q$ in Eq. \eqref{GCA-MEMeAction} as a potential; in that case, the requirement that the solutions be dynamically stable suggests that $\lambda<0$. The parameter $\lambda$ may be interpreted as a vacuum energy for the matter fields, and a negative vacuum energy is expected for matter models motivated by string theory and supersymmetry \cite{Witten2000}. At energy scales close to $\lambda$, the Jordan-frame metric (and $A{_\mu}{^\alpha}$) becomes degenerate, which is compatible with the general expectation that the vacuum energy corresponds to the scale at which the effective spacetime geometry breaks down. On the other hand, the gravitational metric remains well behaved in this limit, with Eq. \eqref{GCA-GEN-GFE-MEMe} reducing to the Einstein field equations for a de Sitter or anti de Sitter vacuum with effective cosmological constant $-\lambda$; this property has been used in \cite{FengCarloni2019} to show that the MEMe model qualitatively exhibits inflationary behavior in the early Universe for $\lambda<0$.

The reader may note that the MEMe model superficially resembles other modified gravity theories that can be interpreted as a modification of the gravitational coupling, such as scalar-tensor theory or disformal theories  \cite{BeltranJimenez2016,Gumrukcuoglu2019a,*Gumrukcuoglu2019b,*DeFelice2019,Bekenstein1993,*Zuma2014,*Kimura2017,*Papadopoulos2018,*Domenech2018, Kaloper2004,*Bettoni2013,*Zumalacarregui2013,*Deruelle2014,*Minamitsuji2014,*Watanabe2015,*Motohashi2016,*Domenech2015a,*Domenech2015b,*Sakstein2015,*Fujita2016,*vandeBruck2017,*Sato2018,*Firouzjahi2018,ClaytonMoffat1999,*ClaytonMoffat2000,*ClaytonMoffat2003,*Moffat2003,*Magueijo2009,*Moffat2016}. Indeed, as pointed out in \cite{FengCarloni2019}, the Jordan metric  $\mathfrak{g}_{\mu \nu}$ may be viewed as a type of vector disformal transformation \cite{Bekenstein1993,*Zuma2014,*Kimura2017,*Papadopoulos2018,*Domenech2018}. However, the difference here is that the MEMe model, being an MMG, does not introduce additional dynamical degrees of freedom; the components of the tensor $A{_\mu}{^\alpha}$ can be expressed directly in terms of the fluid quantities $\rho$, $p$, $u_\mu$ and the metric. As discussed in \cite{Panietal2013}, the addition of an auxiliary field in a gravitational theory will generically produce terms involving derivatives of the energy-momentum tensor in the field equations. While the MEMe model evades this problem in the Einstein frame, the derivatives of $A{_\mu}{^\alpha}$ present in the Jordan-frame equations discussed in the preceding paragraph will by way of Eq. \eqref{GCA-ExpFEs} generate terms containing up to second-order derivatives of $\mathfrak{T}_{\mu \nu}$. The standard PPN formalism is not equipped to handle such terms, and in the following sections, we propose and develop methods for dealing with this obstacle.


%
%
%
%
\section{Newtonian limit of the MEMe model}
\label{sec:Newtonian}

It is helpful to first consider the Newtonian limit of the MEMe model. In doing so, we will assume that $q$ is at most of order one. Such a choice is motivated by the values that we have found for the modulus of $q$ in \cite{FengCarloni2019}. This assumption, combined with the smallness assumption on $\underline{\rho}$ that is made in the Newtonian and post-Newtonian analysis, implies that in our calculation we have at most $q\underline{\rho} \sim O(\epsilon)$, where $\epsilon=1/c^2$.

Our primary aim in this section is to identify and study the Newtonian potential in the MEMe model. We begin by expressing Eq. \eqref{GCA-GEN-GFE-EF} in the form
\begin{equation}\label{GCA-GEN-GFE-EF-2}
R_{\mu \nu}= 8 \pi G \,\left(T_{\mu \nu}-\frac{1}{2}g_{\mu \nu}T\right),
\end{equation}
where $T=g_{\alpha \beta}T^{\alpha \beta}$.

In an appropriately chosen coordinate system (see also Chap. 4 of \cite{Will1993,Will2018} for further discussion), the $(0,0)$ component becomes
\begin{equation}
R_{0 0} \approx 4 \pi G \, T_{0 0},
\end{equation}
where we have used the fact that in the Newtonian limit
\begin{equation}
\frac{T_{i j}}{T_{0 0}} \ll 1.
\end{equation}
From Eq. \eqref{GCA-ExpTmnEffDecompTs}, and taking into account the fact that in our approximation $\det(A)\approx 1$, we obtain
\begin{equation}
T_{0 0}\approx\underline{\rho},
\end{equation}
so that, defining $R_{0 0}=\Delta \Phi_E$ (with $\Delta$ being the Laplacian),
\begin{equation}\label{NLEF}
\Delta \Phi_E= 4 \pi G \underline{\rho}.
\end{equation}
However, from an operational point of view, an accelerometer would measure the Newtonian limit of the Jordan-frame metric $\mathfrak{g}_{\mu \nu}$. Such a potential would be related to $\Phi_E$ by the relation
\begin{equation}\label{NLJF}
\Phi_J = \Phi_E + C \Delta \Phi_E.
\end{equation}
In the case of MEMe, the coefficient $C$ is given by
\begin{equation}
 C=\frac{3 q}{16 \pi G}.
\end{equation}
In order to preserve the traditional notation we will from this point on work in terms of a potential $U$ satisfying an equation of the same form as Eq. \eqref{NLEF}. While it is convenient to work in terms of a potential satisfying Eq. \eqref{NLEF}, one should keep in mind that the physically relevant potential is $\Phi_J$, which we will relate to $U$ as we develop the extended PPN formalism in the next section.

The expression for $\Phi_J$ in Eq. \eqref{NLJF} brings up a potential conceptual difficulty. If $\underline{\rho}$ has a sharp discontinuity, as one might expect at the boundary of a star, the gradient of $\Phi_J$ can be large---a similar difficulty has been identified in the qualitatively similar EiBI theory \cite{PaniSotiriouPRL2012}. However, a large gradient in $\Phi_J$ implies a strong gravitational force, which would lead to a rearrangement of matter. One would expect this gravitational backreaction on the matter distribution to drive the system away from large gradients in $\Phi_J$ (similar arguments \cite{Kim2014} have been made for the corresponding difficulty in EiBI---see also \cite{BeltranJimenezetal2017}).


%
%
%
%
\section{Extended PPN formalism}
\label{sec:PPN}

Naively, one might expect that the PPN formalism applied to generalized coupling theories in the Einstein frame yields a set of PPN parameters which are the same as those of general relativity. In the MEMe model, for instance, the theory is identical to GR if the energy-momentum tensor $T_{\mu \nu}$ as defined in Eq. \eqref{GCA-ExpTmnEffDecomp} has the perfect fluid form. However, as established in \cite{FengCarloni2019}, the Jordan-frame metric is the physically relevant metric, since it is the metric which couples directly to matter. Moreover, the microscopic description of matter is specified by the action of matter fields minimally coupled to the metric in the Jordan frame and thus gives the equation of state of the matter fluid in the Jordan frame. It is therefore appropriate to introduce the PPN potentials and parameters in the Jordan frame. On the other hand, it is more convenient to perform most of the computations in the Einstein frame. Notice that the distinction between the two frames concerns only physical systems in which matter sources play important roles, and therefore it does not concern the correction to e.g. celestial mechanics on Solar System scales.

It may be helpful to provide a brief overview of our procedure, which we first develop for a more general class of modified gravity theories and generalized coupling theories and then apply to the MEMe model. We first attempt to apply the PPN formalism to the Jordan-frame metric, but we find that to avoid higher-order derivatives of the PPN potentials in the field equations, counterterms must be added to the Jordan-frame metric. We then express the Einstein-frame metric in terms of Jordan-frame variables so that we can use the simpler field equation \eqref{GCA-GEN-GFE-EF} in the PPN analysis.

\subsection{Standard PPN formalism}
We follow the conventions of \cite{Will2018} with the post-Newtonian bookkeeping [with the mass density being defined as $\underline{\rho} := \rho (1+\Pi)$]:
\begin{equation}\label{GCA-PPNbookkeeping}
U \sim v^2 \sim p/\rho \sim \Pi \sim \mathcal{O}(\epsilon),
\end{equation}

\noindent so that the velocity components $v^i$ are of order $\mathcal{O}(\epsilon^{1/2})$. It should be mentioned that $v^i$, which are raised components of the three-velocity in the Jordan frame, do not correspond directly to the components of $u_\mu$ but to the raised index four-velocity $\mathfrak{u}^\mu$ in the Jordan frame. Recall that the distinction between $\mathfrak{u}^\mu$ and $u_\mu$ is necessary because there are two metric tensors in generalized coupling models. The components of $\mathfrak{u}^\mu$ have the explicit form
\begin{equation}\label{GCA-FourVelocityDecomp}
\mathfrak{u} = \left( \mathfrak{u}^0, \mathfrak{u}^0 \, \vec{v} \right),
\end{equation}

\noindent where $\vec{v}$ is the coordinate three-velocity of the fluid in the Jordan-frame with components $v^i$. In terms of Jordan frame fluid quantities, one may use Eqs. \eqref{GCA-EinsteinFrame4v} and \eqref{GCA-ExpTmnEffDecomp} to write the source of the gravitational field equation \eqref{GCA-GEN-GFE-EF} as follows:
\begin{equation}\label{Apdx-GCA-ExpTmnEffDecomp}
T_{\mu \nu} = - \varepsilon \left(\tau_1 + \tau_2\right) \, \mathfrak{u}_\mu \, \mathfrak{u}_\nu + \tau_2 \, g_{\mu \nu} .
\end{equation}

Following \cite{Will2018} (and the coordinate conventions therein), we introduce the conserved rest mass density $\rho^*$ which is defined according to the following formula:
\begin{equation}\label{GCA-RestMassDensity}
\rho^* := \sqrt{-\mathfrak{g}} \, \mathfrak{u}^0 \, \rho = |A{_\cdot}{^\cdot}| \sqrt{-g} \, \mathfrak{u}^0 \, \rho.
\end{equation}

\noindent Given $\rho^*$, one may then define the following PPN potentials by the differential relations:\footnote{We point out to the reader that while the PPN formalism in \cite{Will2018} is equivalent to that of \cite{Will1993}, the definitions of the PPN potentials have changed (though the PPN parameters are the same); where the PPN potentials in \cite{Will1993} are defined with respect to $\rho$, the PPN potentials in \cite{Will2018} are defined with respect to $\rho^*$. This change results in a change in the coefficients in front of the potentials $\Phi_1$ and $\Phi_2$ in Eq. \eqref{GCA-metricPPN00} for the metric component $\tilde{\mathfrak{g}}_{00}$.}
\begin{align}
\Delta U &= - 4 \pi G \, \rho^* \label{GCA-PPNpotU}\\
\Delta V_i &= - 4 \pi G \, \rho^* \, v_i \label{GCA-PPNpotV}\\
\Delta W_i &= - 4 \pi G \, \rho^* \, v_i + 2 \partial_i\partial_t U \label{GCA-PPNpotW}\\
\Delta \Phi_1 &= - 4 \pi G \, \rho^* \, v^2 \label{GCA-PPNpot1}\\
\Delta \Phi_2 &= - 4 \pi G \, \rho^* \, U \label{GCA-PPNpot2}\\
\Delta \Phi_3 &= - 4 \pi G \, \rho^* \, \Pi \label{GCA-PPNpot3}\\
\Delta \Phi_4 &= - 4 \pi G \, p , \label{GCA-PPNpot4}
\end{align}

\noindent and the following potentials by integral relations:
\begin{align}
\Phi_6
&= G \int {\rho^*}^\prime \frac{\left[\vec{v}\cdot(\vec{x}-\vec{x}^\prime)\right]^2}{|\vec{x}-\vec{x}^\prime|^3} d^3 x^\prime \label{GCA-PPNpot6}\\
\Phi_W
&= G \int \int {\rho^*}^\prime {\rho^*}^{\prime \prime} \frac{\vec{x}-\vec{x}^\prime}{|\vec{x}-\vec{x}^\prime|^3} \cdot \left[\frac{\vec{x}^\prime - x^{\prime \prime}}{|\vec{x}^\prime - x^{\prime \prime}|}\right] d^3 x^\prime d^3 x^{\prime \prime} \label{GCA-PPNpotPhiW} \\
&\qquad - \int \int {\rho^*}^\prime {\rho^*}^{\prime \prime} \frac{\vec{x}-\vec{x}^\prime}{|\vec{x}-\vec{x}^\prime|^3} \cdot \left[\frac{\vec{x} - x^{\prime \prime}}{|\vec{x}^\prime - x^{\prime \prime}|}\right] d^3 x^\prime d^3 x^{\prime \prime} . \nonumber
\end{align}

\noindent In the standard PPN formalism, the metric tensor is expanded as follows:
\begin{align}
\underline{g}_{00} =
& - 1 + 2 U -2 \beta U^2 + (2\gamma + 1 + \alpha_3 + \zeta_1 - 2 \xi) \Phi_1 \nonumber \\
& + 2(1 - 2\beta +\zeta_2 +\xi) \Phi_2 + 2(1 + \zeta_3) \Phi_3 \nonumber \\
& + 2(3\gamma + 3 \zeta_4 -2 \xi)\Phi_4 - (\zeta_1 - 2\xi) \Phi_6 - 2 \xi \Phi_W \label{GCA-GENmetricPPN00} \\
\underline{g}_{0j} =
& - \tfrac{1}{2} (4\gamma + 3 + \alpha_1 - \alpha_2 + \zeta_1 -2 \xi) V_j \nonumber \\
& - \tfrac{1}{2} (1 + \alpha_2 - \zeta_1 + 2 \xi) W_j , \label{GCA-GENmetricPPN0j} \\
\underline{g}_{ij} =& (1 + 2 \gamma U) \delta_{ij} . \label{GCA-GENmetricPPNij}
\end{align}

\noindent The metric is inserted into the field equations and expanded to PPN order $\mathcal{O}(\epsilon)$; one then matches terms proportional to each of the potentials in Eqs. \eqref{GCA-PPNpotU}-\eqref{GCA-PPNpotPhiW} to obtain the PPN coefficients.

\subsection{Extended PPN formalism}
The procedure outlined in the preceding section does not suffice for certain classes of modified gravity theories. For instance, one might imagine in four dimensions a rather general theory of the form (use of the Cayley-Hamilton theorem has been employed on the rhs):
\begin{align}
R{^\mu}{_\nu} + e{^\mu}{_\nu} = & ~ A_1(\mathfrak{T}{^\cdot}{_\cdot}) \, \mathfrak{T}{^\mu}{_\nu} + A_2(\mathfrak{T}{^\cdot}{_\cdot}) \, \mathfrak{T}{^\mu}{_\alpha} \mathfrak{T}{^\alpha}{_\nu} \nonumber \\
& + A_3(\mathfrak{T}{^\cdot}{_\cdot}) \, \mathfrak{T}{^\mu}{_\alpha} \mathfrak{T}{^\alpha}{_\beta} \mathfrak{T}{^\beta}{_\nu} + B(\mathfrak{T}{^\cdot}{_\cdot}) \, \delta{^\mu}{_\nu},
\label{GCA-AGeneralCouplingModel}
\end{align}

\noindent where $e{^\mu}{_\nu}$ contains additional geometric or gravitational terms, $\mathfrak{T}{^\mu}{_\nu}$ is the energy-momentum tensor, and $A_i(\mathfrak{T}{^\cdot}{_\cdot})$ and $B(\mathfrak{T}{^\cdot}{_\cdot})$ are scalar functions that are polynomials in scalar invariants of $\mathfrak{T}{^\mu}{_\nu}$ up to third order. Examples of such a theory include the EiBI  \cite{BanadosFerreira2010} or the braneworld model of \cite{Shiromizu1999}. We also note that Eq. \eqref{GCA-AGeneralCouplingModel} is also a subcase of the gravitational field equation given in \cite{Panietal2013}.

We consider a class of type-I MMG theories in which the source terms in the Einstein frame can be written exclusively in terms of the energy-momentum tensor so that $e{^\mu}{_\nu}=0$. Expanding the rhs of Eq. \eqref{GCA-AGeneralCouplingModel} to post-Newtonian order, one has a term proportional to $\rho^2$; however, the PPN expression for the Ricci tensor does not contain any term that can absorb such a term. One may remedy this by adding a term to $\underline{g}_{00}$ \eqref{GCA-GENmetricPPN00} of the form\footnote{Here, we follow the conventions of \cite{Will2018}. If one wishes to use those of \cite{Will1993}, one should instead add a term of the form $\nu \Psi^{\circ}$, where $\Psi^{\circ}$ is defined similarly to $\Psi$ but with $\rho$ instead of $\rho^*$.} $2 \nu \Psi$, where $\Psi$ is a $\mathcal{O}(\epsilon^2)$ potential defined by the following:
\begin{equation}\label{GCA-NewPPNPotential}
\Delta \Psi := - 4 \pi G^2 \, {\rho^*} \rho = - 4 \pi G^2 \, {\rho^*}^2 + O(\epsilon^3).
\end{equation}

\noindent We note here that unlike the standard PPN potentials, this additional potential $\Psi$ is dimensionful---since the metric components must be dimensionless, it follows that the associated parameter $\nu$ must also be dimensionful. We attribute this to the fact that the coefficient for the $\rho^2$ term which appears in the PPN expansion of \eqref{GCA-AGeneralCouplingModel} introduces an additional scale into the theory. Later, we shall see this explicitly when applying this extended PPN formalism to the MEMe model.

We now turn to the case of generalized coupling theories as described by Eqs. \eqref{GCA-GENAction} and \eqref{GCA-AuxiliarymetricGC}. For an appropriate choice of reference frame, the extended PPN metric for the Jordan-frame metric would take the form (note the addition of the term $2 \nu \Psi$ in $\tilde{\mathfrak{g}}_{00}$)
\begin{align}
\tilde{\mathfrak{g}}_{00} =
& - 1 + 2 U -2 \beta U^2 + (2\gamma + 1 + \alpha_3 + \zeta_1 - 2 \xi) \Phi_1 \nonumber \\
& + 2(1 - 2\beta +\zeta_2 +\xi) \Phi_2 + 2(1 + \zeta_3) \Phi_3 \nonumber \\
& + 2(3\gamma + 3 \zeta_4 -2 \xi)\Phi_4 - (\zeta_1 - 2\xi) \Phi_6 - 2 \xi \Phi_W \nonumber \\
& + 2 \nu \Psi, \label{GCA-metricPPN00} \\
\tilde{\mathfrak{g}}_{0j} =
& - \tfrac{1}{2} (4\gamma + 3 + \alpha_1 - \alpha_2 + \zeta_1 -2 \xi) V_j \nonumber \\
& - \tfrac{1}{2} (1 + \alpha_2 - \zeta_1 + \xi) W_j , \label{GCA-metricPPN0j} \\
\tilde{\mathfrak{g}}_{ij} =& (1 + 2 \gamma U) \delta_{ij} . \label{GCA-metricPPNij}
\end{align}

\noindent However, one still encounters a difficulty when attempting to apply the standard PPN analysis to Eq. \eqref{GCA-GEN-GFE-MEMe}. As discussed earlier, the gravitational field equations in the Jordan frame will contain up to second-order derivatives of $\mathfrak{T}_{\mu \nu}$. It follows that the direct application of the PPN form to the Jordan-frame metric will introduce terms involving second derivatives of the fluid potentials and four-velocity, but the standard PPN formalism and the extended formalism encapsulated in Eqs. (\ref{GCA-metricPPN00})--(\ref{GCA-metricPPNij}) are incapable of absorbing these terms. To see this, consider the following expression for the Einstein-frame metric $g_{\mu \nu}$:
\begin{equation}\label{GCA-MetricJordanFrame}
g_{\mu \nu} = \Xi^{-1} \, \bar{A}{^\alpha}{_\mu} \, \bar{A}{^\beta}{_\nu} \, \mathfrak{g}_{\alpha \beta} ,
\end{equation}

\noindent From Eq. \eqref{GCA-GEN-AFE}, the tensor $\bar{A}{^\alpha}{_\mu}$ and the factor $\Xi=\Xi(A{_\cdot}{^\cdot})$ depend on $\rho^*$, $\Pi$ and $p$, and we assume $\mathfrak{g}_{\alpha \beta}$ takes the usual PPN form given in Eqs. \eqref{GCA-metricPPN00}--\eqref{GCA-metricPPNij}.
Upon expanding the Ricci tensor for $g_{\mu \nu}$ as given by \eqref{GCA-MetricJordanFrame} into Eq. \eqref{GCA-GEN-GFE-MEMe}, one will obtain terms containing derivatives of $\rho^*$, $\Pi$ and $p$, which cannot be absorbed by remaining terms in Eq. \eqref{GCA-AGeneralCouplingModel} if $e{^\mu}{_\nu}=0$.\footnote{One might suppose that $e{^\mu}{_\nu}$ contains terms with derivatives of $\rho^*$, $\Pi$ and $p$, which can cancel the additional terms introduced by $\Xi$ and $\bar{A}{^\alpha}{_\mu}$. Derivatives of $\rho^*$, $\Pi$ and $p$ correspond to higher-order ($>2$) derivatives of the potentials, which correspond to higher-order derivatives of the metric---one then has a higher-order theory of gravity, which (excluding frame-dependent theories like Ho\ifmmode \check{r}\else \v{r}\fi{}ava-Lifshitz gravity \cite{Horava2009} and a certain class of type-II MMG theories \cite{Aoki:2020lig,Aoki:2020iwm,Aoki:2020ila,Yao:2020tur}) generically suffers from Ostrogradskian instability \cite{Ostrogradsky1850,Woodard2015}.}

To eliminate these additional terms, we can add counterterms to the metric components $\tilde{\mathfrak{g}}_{\mu \nu}$ given in Eqs. \eqref{GCA-metricPPN00}--\eqref{GCA-metricPPNij} and then choose coefficients such that
Eq. \eqref{GCA-MetricJordanFrame} does not contain the quantities $\rho^*$, $\Pi$ and $p$. In general, the counterterms take the following form:
\begin{align}
{\mathfrak{g}}_{00} =
& \tilde{\mathfrak{g}}_{00} + c_0 \Delta U + c_1 \Delta \Phi_1 + c_2 \Delta \Phi_2 + c_3 \Delta \Phi_3 \nonumber\\
& + c_4 \Delta \Phi_4 + c_\Psi \Delta \Psi + c_w \Delta \Phi_W , \label{GCA-JmetricPPN00}\\
{\mathfrak{g}}_{0j} =
& \tilde{\mathfrak{g}}_{0j} + d_V \Delta V_j + d_W \Delta W_j, \label{GCA-JmetricPPN0j} \\
{\mathfrak{g}}_{ij} =
    & \tilde{\mathfrak{g}}_{ij} + e_0 \Delta U \delta_{ij}
\label{GCA-JmetricPPNij}
\end{align}

\noindent where we restrict to terms of order $\mathfrak{g}_{00} \sim \mathcal{O}(\epsilon^2)$, $\mathfrak{g}_{0j} \sim \mathcal{O}(\epsilon^{3/2})$, and $\mathfrak{g}_{ij} \sim \mathcal{O}(\epsilon)$. At this stage, one may collect terms of order $\epsilon$ in ${\mathfrak{g}}_{00}$ which yields the Newtonian potential in the Jordan frame:
\begin{equation}
\Phi_J = U + (c_0/2) \Delta U,
\end{equation}

\noindent consistently with what was obtained in \eqref{NLJF}. We then choose the coefficients $c_{0-4,\Psi,w}$, $d_{V,W}$ and $e_0$ so that the Einstein-frame metric takes the desired form:
\begin{align}
g_{00} =
& - 1 + 2 U - 2 \beta U^2 + (2\gamma + 1 + \alpha_3 + \zeta_1 - 2 \xi) \Phi_1 \nonumber \\
& + 2(1 - 2\beta +\zeta_2 +\xi) \Phi_2 + 2(1 + \zeta_3) \Phi_3 \nonumber \\
& + 2(3\gamma + 3 \zeta_4 -2 \xi)\Phi_4 - (\zeta_1 - 2\xi) \Phi_6 - 2 \xi \Phi_W \nonumber \\
& + 2 \nu \Psi, \label{GCA-EFmetricPPN00} \\
g_{0j} =
& - \tfrac{1}{2} (4\gamma + 3 + \alpha_1 - \alpha_2 + \zeta_1 -2 \xi) V_j \nonumber \\
& - \tfrac{1}{2} (1 + \alpha_2 - \zeta_1 + 2 \xi) W_j , \label{GCA-EFmetricPPN0j} \\
g_{ij} =& (1 + 2 \gamma U) \delta_{ij} . \label{GCA-EFmetricPPNij}
\end{align}

\noindent where again we restrict to terms of order
$g_{00} \sim \mathcal{O}(\epsilon^2)$, $g_{0j} \sim \mathcal{O}(\epsilon^{3/2})$, and $g_{ij} \sim \mathcal{O}(\epsilon)$. The reader should keep in mind here that all of the potentials in this expression are those appearing in Eqs. \eqref{GCA-metricPPN00}--\eqref{GCA-metricPPNij}, which are defined with respect to Jordan-frame fluid quantities. Therefore this expression is not a PPN expansion of the Einstein-frame metric---rather, one should think of Eqs. \eqref{GCA-EFmetricPPN00}--\eqref{GCA-EFmetricPPNij} as the Einstein-frame metric expressed in terms of (PPN expanded) Jordan-frame quantities.

It is worth mentioning at this point that to post-Newtonian order, the metric $\mathfrak{g}_{\mu \nu}$ retains the form expected for the PPN gauge in the sense that the spatial components ${\mathfrak{g}}_{ij}$ do not acquire cross terms. It should also be mentioned that we are in fact working in a PPN gauge since $\tilde{\mathfrak{g}}_{ij}$ is diagonal and depends strictly on the potentials \eqref{GCA-PPNpotU}--\eqref{GCA-PPNpotPhiW}---from Ch. 4 of \cite{Will2018}, we expect that a non-PPN gauge will introduce an additional potential. To clarify, one first chooses the gauge in which $\tilde{\mathfrak{g}}_{\mu \nu}$ has the form given in Eqs. \eqref{GCA-metricPPN00}--\eqref{GCA-metricPPNij}; after the gauge is chosen, the set of counterterms in Eqs. \eqref{GCA-JmetricPPN00}--\eqref{GCA-JmetricPPNij} for $\mathfrak{g}_{\mu \nu}$ is sufficient to characterize the PPN expansion.

The proposed modification to the PPN parameterization has been motivated by necessity; without these modifications, one cannot apply the PPN formalism to a class of type-I MMGs and GCTs whose equations of motion can be written in the form of Eq. \eqref{GCA-AGeneralCouplingModel} (with $e{^\mu}{_\nu}=0$), including the MEMe model. Though we have provided here a preliminary discussion regarding the theoretical interpretation for the new potential $\Psi$, it is perhaps appropriate to also understand the physical interpretation of $\Psi$ and the counterterms in a phenomenological context. We will attempt to address this point in later sections by studying the net effect of these quantities on some post-Newtonian systems in the MEMe model.

\subsection{MEMe model coefficients}
We now apply the extended PPN formalism described above to the MEMe model. First, we note that in Eq. \eqref{GCA-MetricJordanFrame}, $\Xi=1$ for the MEMe model [compare Eqs. \eqref{GCA-AuxiliarymetricGC} and \eqref{GCA-JordanMetric} and recall that $A=4$ on shell]. We then demand that the Einstein-frame metric $g_{\mu \nu}$ has the form given in Eqs. \eqref{GCA-EFmetricPPN00}--\eqref{GCA-EFmetricPPNij}, and upon comparison with Eq. \eqref{GCA-MetricJordanFrame} for the MEMe model, one obtains the following values for the coefficients of the counterterms:
\begin{align}
  & c_0 = \frac{3 q}{8 \pi G}, \quad c_1 = \frac{5 q}{16 \pi G}, \quad c_2 = -\frac{3 (3 \gamma +2) q}{8 \pi  G}, \nonumber \\
  & c_3 = \frac{3 q}{8 \pi  G}, \quad c_4 = \frac{3 q}{8 \pi  G} , \quad c_\Psi = \frac{21 q^2}{64 \pi  G^2}, \label{GCA-Coefficients49a}
\end{align}

\begin{equation}\label{GCA-Coefficients50}
    e_0 = \frac{q}{8 \pi  G}, \quad
    c_w = 0, \quad
    d_V = -\frac{q}{2 \pi  G}, \quad
    d_W = 0 .
\end{equation}

The expression for the Einstein-frame metric ${g}_{\mu \nu}$ in Eqs. \eqref{GCA-EFmetricPPN00}--\eqref{GCA-EFmetricPPNij} is then substituted into Eq.\eqref{GCA-GEN-GFE-EF}, and we find that all of the standard PPN parameters are exactly the same as that of general relativity ($\gamma=\beta=1$, all others zero). However, the new parameter $\nu$, which has the value $\nu = 0$ in general relativity, has the following value in the MEMe model:
\begin{equation}\label{GCA-NewPPNparamMEMe}
\nu = \frac{3 q}{2 G}.
\end{equation}

\noindent As anticipated by our remarks in the preceding section, the parameter $\nu$ corresponds to the scale $q=1/\lambda$ that appears in the MEMe model.


%
%

%
%

\section{Monopole term for PPN potentials}
\label{sec:monopole}

\subsection{General analysis}
 We will now investigate the physical effects of the modification of the PPN monopole term associated with $\Psi$.  We begin by assuming that the matter distribution is compact and static (so that $v^i=0$) and consider what happens outside the matter distribution. One may then define an effective gravitational potential in the following manner:
\begin{equation}\label{GCA-PotentialDef}
\Phi :=
\frac{1}{2}\left( 1 + g_{00} + 2 \, \beta \, U^2\right).
\end{equation}

\noindent Outside a matter distribution, the counterterms vanish---recall that outside of a matter distribution, the Einstein- and Jordan-frame metrics coincide. For a theory with no preferred location effects ($\xi=0$), the effective gravitational potential takes the form (we set $v^i=0$ so that $\Phi_1=\Phi_6=0$)
\begin{equation}\label{GCA-Potential}
\begin{aligned}
\Phi = & ~ U + 2 \, \beta_2 \, \Phi_2 + \beta_3 \, \Phi_3 + 3 \, \beta_4 \, \Phi_4 + \nu \, \Psi.
\end{aligned}
\end{equation}

\noindent where (following the reasoning in Chap. 40 of \cite{MTW}):
\begin{equation}\label{GCA-Betas}
\begin{aligned}
\beta_2 &:= \frac{1}{2} \left(1 - 2\beta + \zeta_2 \right) \\
\beta_3 &:= 1 + \zeta_3 \\
\beta_4 &:= \gamma + \zeta_4.
\end{aligned}
\end{equation}

\noindent Note that up to an overall factor of $2$, $\Phi$ consists of all terms in $\mathfrak{g}_{00}$ such that $\Delta \Phi$ can be written as an algebraic function of $\rho$, $\Pi$, $p$, and $U$ up to fourth order in $\epsilon$. We consider the case where the gravitational theory is fully conservative, with the parameter choices $\alpha_1=\alpha_2=\alpha_3=\zeta_1=\zeta_2=\zeta_3=\zeta_4=0$ (in addition to $\xi=0$); one has $\beta_2=(1 - 2 \beta)/2$, $\beta_3=1$, and $\beta_4=\gamma$.

We now consider the multipole expansion for the Newtonian potential:
\begin{equation}\label{GCA-NewtonianPotential}
\Phi(x) = \int \frac{G \, \rho_e(x^{\prime})}{|\textbf{x}-\textbf{x}^{\prime}|} \, d^3 x^{\prime},
\end{equation}

\noindent where $\rho_{e}$ is an effective energy density given by
\begin{equation}\label{GCA-rhoeff}
\rho_{e} = \rho^* \biggl[ 1 + 2\beta_2 \, U + \beta_3 \, \Pi + 3 \beta_4 \, {p}/{\rho^*} + \nu \, G \, \rho \biggr].
\end{equation}

\noindent The monopole moment is given by
\begin{equation}\label{GCA-NewtonianPotentialMonopole}
\Phi(x) = \frac{G \, M}{r} + O(r^{-2}),
\end{equation}

\noindent where
\begin{equation}\label{GCA-massdef}
M := \int \rho_{e}(x^{\prime}) \, d^3 x^{\prime}.
\end{equation}

\noindent The definition given in Eqs. \eqref{GCA-rhoeff} and \eqref{GCA-massdef} is motivated by Eq. (40.4) in \cite{MTW}; it is in fact identical in the limit $\nu \rightarrow 0$.

For the case of a stationary spherical mass $W_i=V_i=0$, $\mathcal{A}=\Phi_1=0$. Making use of the fact that $\rho_e^2={\rho^*}^2+O(\epsilon^3)$, and keeping only the monopole terms, the metric to post-Newtonian order is [cf. Eq. (40.3) of  \cite{MTW}]:
\begin{align}
\mathfrak{g}_{00} =
& - 1 + \frac{2 \, G \, M}{r} - \frac{2 \, \beta \, G^2 \, M^2 }{r^2} \label{GCA-metricPPN002}\\
\mathfrak{g}_{0j} =& ~ 0 , \label{GCA-metricPPN0j2} \\
\mathfrak{g}_{ij} =& \left[1 + \frac{2 \, \gamma \, G \, M}{r} \right] \delta_{ij} . \label{GCA-metricPPNij2}
\end{align}

\noindent It follows that for a spherically symmetric matter distribution, the additional PPN potential can be absorbed into the mass, as one might have expected. This suggests that outside of a spherically symmetric matter distribution, the effects of the additional potential $\Psi$ cannot be disentangled from the other potentials.

To distinguish the effects of the potential $\Psi$ and parameter $\nu$, one should consider the internal structure of the source. In particular, if one has a detailed model for the source itself, it may be possible to disentangle the effects of the parameter $\nu$ from the total mass of a spherical source. To see how one might distinguish the effects of an additional potential $\Psi$, we consider a given matter distribution and split the mass $M$ into two parts: one which depends on the original PPN parameters and one which depends on the new parameter $\nu$. Defining the potential
\begin{equation}\label{GCA-PotentialTilde}
\begin{aligned}
\bar{\Phi} := \Phi-\nu \, \Psi
\end{aligned}
\end{equation}

\noindent and defining $\bar{\rho}_e := \rho_e - \nu \, G \, \rho^*\rho$, one has the result
\begin{equation}\label{GCA-NewtonianPotentialMonopoleTilde}
\bar{\Phi}(x) = \frac{G \, \bar{M}}{r} + O(r^{-2}),
\end{equation}

\noindent where the mass defined with respect to the original PPN potentials takes the form
\begin{equation}\label{GCA-massdefbar}
\bar{M} := \int \bar{\rho}_{e}(x^{\prime}) \, d^3 x^{\prime}.
\end{equation}

\noindent Now we consider the standard multipole expansion for the new PPN potential:
\begin{equation}\label{GCA-NewPPNPotentialbar}
\Psi(x) = \int \frac{G^2 \, \rho^*(x^{\prime})\rho(x^{\prime})}{|\textbf{x}-\textbf{x}^{\prime}|} d^3 x^{\prime}.
\end{equation}

\noindent Now $\rho^*\rho=\bar{\rho}_e^2+O(\epsilon^3)$. The monopole moment is given by
\begin{equation}\label{GCA-NewPPNPotentialMonopolebar}
\Psi(x) = \frac{G^2 \, \mu^2}{r} + O(r^{-2}),
\end{equation}

\noindent where
\begin{equation}\label{GCA-mudef}
\mu^2 := \int \rho^*(x^{\prime})\rho(x^{\prime}) \, d^3 x^{\prime} = \int \bar{\rho}_e(x^{\prime})^2 \, d^3 x^{\prime} + O(\epsilon^3).
\end{equation}

\noindent The relationship between $\bar{M}$ and $\mu^2$ is sensitive to the internal structure of the source. For instance, if one considers the following Gaussian profile for $\bar{\rho}_e$:
\begin{equation}\label{GCA-GaussianProfile}
\bar{\rho}_e(x) = \frac{\bar{M}}{\left(\sqrt{2 \pi} \, \sigma\right)^3} \exp\left[-\frac{r^2}{2 \sigma^2}\right],
\end{equation}

\noindent then one has for $\mu^2$
\begin{equation}\label{GCA-muexp}
\mu^2 = \frac{\bar{M}^2}{8 \, \pi^{3/2} \, \sigma^3}.
\end{equation}

\noindent Note that $\mu^2$ depends on the size $\sigma$ for the source. Motivated by the Gaussian expression, one can use Eq. \eqref{GCA-muexp} as a parameterization for the internal structure of the source, with $\sigma$ being a parameter which represents a characteristic length scale for the source. It follows that\footnote{We note that a $\sigma$-dependent shift in the mass was seen in a different model obtained from considering quantum corrections to the gravitational potential to post-Newtonian order---see Eq. (2.74) of \cite{Casadio2017} and also the approach in \cite{Casadio2018}.}
\begin{equation}\label{GCA-Massexpand}
M = \bar{M} + \nu \, \frac{G \, \bar{M} \, \bar{\rho}_C}{6 \, \sqrt{\pi}},
\end{equation}

\noindent where $\bar{\rho}_C := 3 \bar{M}/ 4 \pi \sigma^3$ is the compactness of the source. Given some matter distribution, the mass $\bar{M}$ is the post-Newtonian mass in the GR limit $\nu \rightarrow 0$ for the MEMe parameter choice. 

\subsection{MEMe model analysis}
It should be mentioned that this dependence on the compactness is only apparent when a detailed description of matter is taken into account. Since MEMe coincides with GR outside matter sources, the inertial mass outside the source is equivalent to the gravitating mass $M$. It follows that one can only compute the difference between the GR value $\bar{M}$ and the MEMe value $M$ when computing the gravitating mass directly from the density. To understand this difference, consider lowering a particle with a small mass $m$ into a matter distribution satisfying the distribution in \eqref{GCA-GaussianProfile}. We consider this process in the Einstein frame since the gravitating mass $M$ in the MEMe model is defined in this frame. The gravitational binding energy between the particle and the matter distribution is given by $m \left[\Phi - \beta U^2\right] = m \left[\bar{\Phi} - \beta U^2 + \nu \Psi\right]$, where
\begin{equation}\label{GCA-GaussianProfileGenPsi}
\Psi(x) = \frac{G^2 \, \bar{M} \, \bar{\rho}_C}{6 \, \sqrt{\pi}} \frac{\text{erf}\left(\frac{r}{\sigma }\right)}{r}.
\end{equation}

\noindent As discussed in \cite{FengCarloni2019}, a stability argument suggests that $q<0$, which in turn suggests $\nu<0$. Since $\Psi(x)>0$, the gravitational binding energy of the particle within a matter distribution is decreased in the MEMe model compared to GR. This result indicates that in the MEMe model, the gravitating mass of an object outside matter sources is less than the sum of its parts due to a weakening of the gravitational binding energy. If the inertial mass and the gravitating mass of an object in a vacuum are the same, then one may then place constraints on the parameter $\nu$ by measuring the mass of an object, disassembling it into its constituent parts, and measuring the mass of the individual components.

\subsection{Constraints on the MEMe model}
One can in principle place a constraint on the parameter $\nu$ without requiring the equivalence of inertial and gravitating masses. To see this, first note that one can interpret Eq. \eqref{GCA-Massexpand} as resulting from a dependence in the effective gravitational constant on the compactness $\bar{\rho}_C$ of the source. For a source mass $\bar{M}$ and the Gaussian profile one has the following expression for the effective gravitational constant:
\begin{equation}\label{GCA-EffectiveGravConstant}
\begin{aligned}
G_{eff} &= G_0 \left[1 +  \frac{\nu \, G_0 \, \bar{\rho}_C}{6 \, \sqrt{\pi}}\right] .
\end{aligned}
\end{equation}
Recent experiments \cite{Lietal2018} with spherical stainless steel (SS 316) source masses, which have a density of $\sim 7.87 \times 10^3 ~ \text{Kg}/\text{m}^3$, constrain Newton's constant to a fractional uncertainty of about $3 \times 10^{-5}$. While the experiment in \cite{Lietal2018} alone cannot place a constraint on $\nu$, one might imagine a variation of the experiment in which the spherical source masses can be disassembled into thick spherical shells. If the same experiment is performed for each shell individually and then again for the fully reassembled source mass, one can search for differences in the effective gravitational constant---such differences are evident of a weakening or strengthening of the gravitational binding energy when masses are brought together. Assuming that fractional uncertainties similar to those of \cite{Lietal2018} can be achieved, one can in principle constrain $\nu$ up to a value on the order of $\nu \sim 10^{-7} ~ \text{m}^3/\text{Kg}$, or $10^{-24} ~ \text{m}^3/\text{J}$, in units of inverse energy. This in turn can place a strong constraint on $q$:
\begin{equation}\label{GCA-qConstraint}
|q| \lessapprox 10^{-24} ~  \text{m}^3/\text{J}.
\end{equation}

\noindent which is $10$ orders of magnitude stronger than the speed of light constraint ($|q| < 2 \times 10^{-14} ~  \text{m}^3/\text{J}$) in \cite{FengCarloni2019}, though still 12 orders of magnitude weaker than scales corresponding to the inverse of the highest energy densities ($\sim 14 \text{GeV}/\text{fm}^3 \approx 2.2 \times 10^{36} \text{J}/\text{m}^3$) probed in accelerator experiments to date \cite{Pasechnik2017,Mitchell2016}, and 26 orders of magnitude weaker than that from a TeV-scale breakdown.


%
%

%
%

\section{Laplacian counterterms and orbits}
\label{sec:counterterms}

\subsection{Circular orbits for conservative theories}
We focus now on the effect of the Laplacian counterterms in the modified PPN metric (\ref{GCA-JmetricPPN00})--(\ref{GCA-JmetricPPNij}) on circular geodesics in the post-Newtonian limit. For simplicity, we assume that matter sources are spherically symmetric and stationary, so that $V_i=0$, $W_i=0$, $\Phi_1=0$, and $\Phi_6=0$. We also consider a conservative theory, which corresponds to the choice
$\alpha_1=\alpha_2=\alpha_3=\zeta_1=\zeta_2=\zeta_3=\zeta_4=0$ in the original PPN analysis of \cite{Will2018}. The line element then has the form ($d\Omega^2$ being the line element on the unit two-sphere):
\begin{equation}\label{WAC-SphericalLineElem}
ds^2 = f \, dt^2 + h \left( {dr^2} + r^2 \, d\Omega^2 \right).
\end{equation}

\noindent To simplify the analysis, we neglect internal energy density and internal pressure. The functions $f$ and $h$ take the following forms:
\begin{align}
f
=& - 1 + 2 U - 2 U^2 + c_0 \Delta U - 2 \Phi_2 + c_2 \Delta \Phi_2 \nonumber \\
& + 2 \nu \Psi + c_\Psi \Delta \Psi , \label{WAC-isof}\\
h
=& 1 + 2 U  - e_0 \Delta U. \label{WAC-isoh}
\end{align}

\noindent For a spherically symmetric matter distribution, one can obtain solutions for the potentials by directly integrating a Poisson equation of the form $\Delta \psi = - 4 \pi \, G \, \rho_s$, which in spherical symmetry may be written explicitly:
\begin{equation}\label{WAC-PoissonEquationSpherical}
\frac{1}{r^2} \frac{\partial }{\partial r}\left(r^2 \frac{\partial \psi(r)}{\partial r}\right) = - 4 \pi \, G \, \rho_s(r)
\end{equation}

\noindent where $\rho_s$ is a source function. This can be integrated to obtain the solution
\begin{equation}\label{WAC-PoissonEquationSphericalSolution}
\psi(r) = C_1 + \int_{r_0}^r \frac{1}{{y}^2}\left[C_2 - {4 \pi G \int_{y_0}^{{y}} \rho_s ({y^\prime}) \, {y^\prime}^2  \, d{y^\prime}}\right]d{y}.
\end{equation}

Given a Jordan-frame geodesic $x^\mu(\tau)$ parameterized by proper time $\tau$, one has the following conserved quantities:
\begin{equation}\label{WAC-ConservedQuantities}
\begin{aligned}
e &= \mathfrak{g}_{\mu 0} \frac{dx^\mu}{d\tau} = f \, \frac{dt}{d\tau} , \\
l &= \mathfrak{g}_{\mu 3} \frac{dx^\mu}{d\tau} = r^2 \, h \, \frac{d\phi}{d\tau} .
\end{aligned}
\end{equation}

\noindent From the unit norm condition for the four-velocity, one can show that the specific energy $e$ must have the form:
\begin{equation}\label{WAC-SpecificEnergy}
e^2 = - f \, h \left(\frac{dr}{d\tau}\right)^2 - \frac{f}{r^2 \, h} l^2 - f.
\end{equation}

\noindent The effective potential may be obtained by considering the turning point (${dr}/{d\tau}=0$) expression for $e^2$:
\begin{equation}\label{WAC-EffectivePotential}
V_{eff} = - f \left[\frac{l^2}{r^2 \, h} + 1 \right].
\end{equation}

We now consider circular orbits and assume spherical symmetry [$f=f(r)$, $h=h(r)$]; circular orbits lie at the minima of the effective potential and are given by the condition $V_{eff}^\prime (r)=0$. One can solve $V_{eff}^\prime (r)=0$ for the specific angular momentum $l$ to obtain
\begin{equation}\label{WAC-Angmom}
l = r \, h \sqrt{\frac{r \, f'}{f \left(r \, h' + 2 \,h\right) - r \, h \, f'}},
\end{equation}

\noindent and a comparison with Eq. \eqref{WAC-ConservedQuantities} yields the proper tangential velocity:
\begin{equation}\label{WAC-SpecificEnergyCircularOrbit}
r \frac{d\phi}{d\tau} = \frac{l}{r \, h(r)} .
\end{equation}

\noindent From the line element Eq. \eqref{WAC-SphericalLineElem}, one has $dt/d\tau = \sqrt{- f(r) - h(r) \, v^2 }$, which yields the tangential coordinate velocity:
\begin{equation}\label{WAC-SpecificEnergyCircularOrbit3}
v(r) \equiv r \frac{d\phi}{d t} = \sqrt{\frac{- r f'(r)}{r h'(r) + 2 h(r)}}.
\end{equation}

\subsection{Circular orbits in the MEMe model}
\begin{figure}[htbp]

(a)

\vspace{-0.25cm}

\includegraphics[width=0.48 \textwidth]{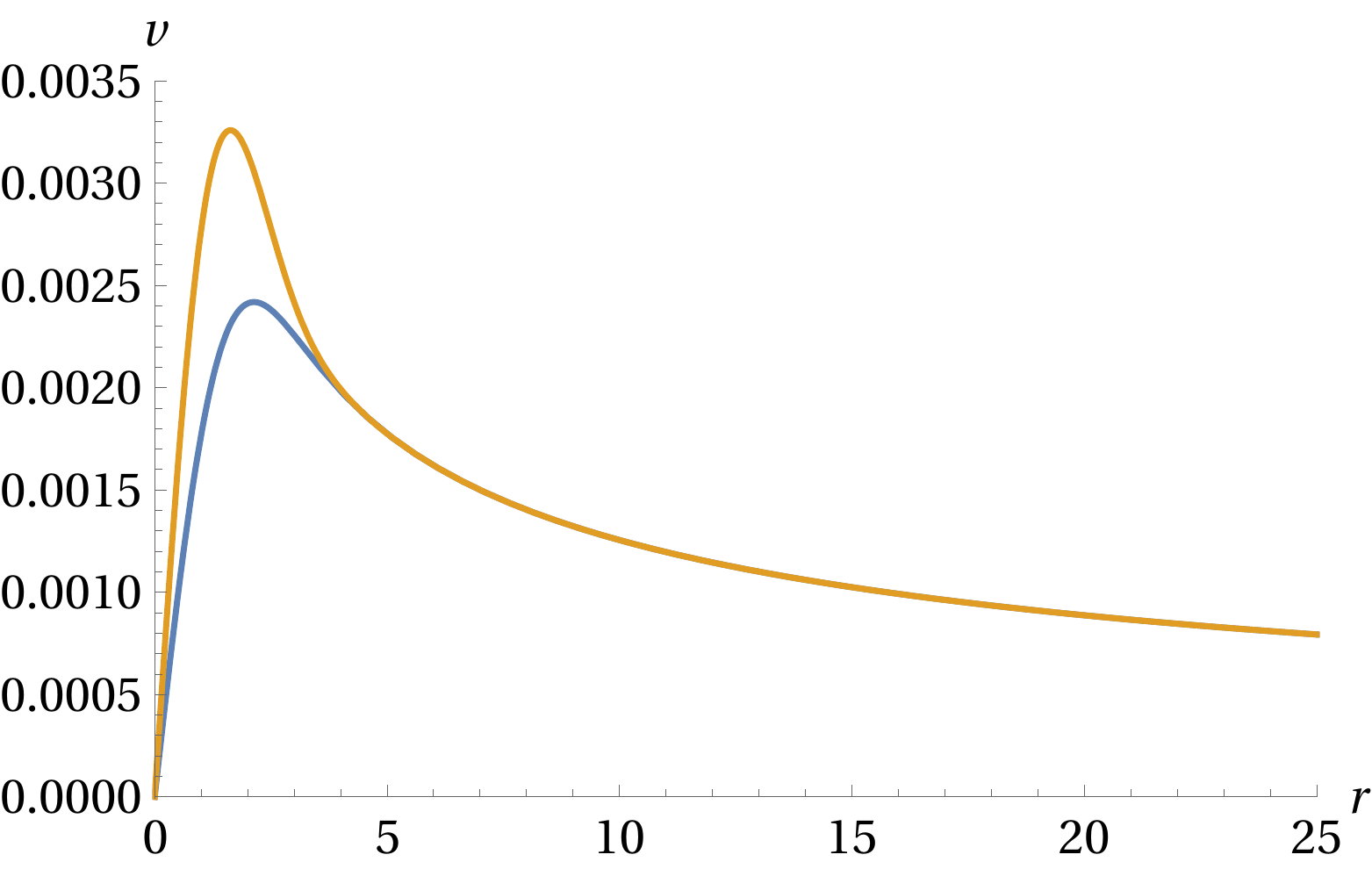}

\vspace{0.25cm}

(b)

\vspace{-0.25cm}

\includegraphics[width=0.48 \textwidth]{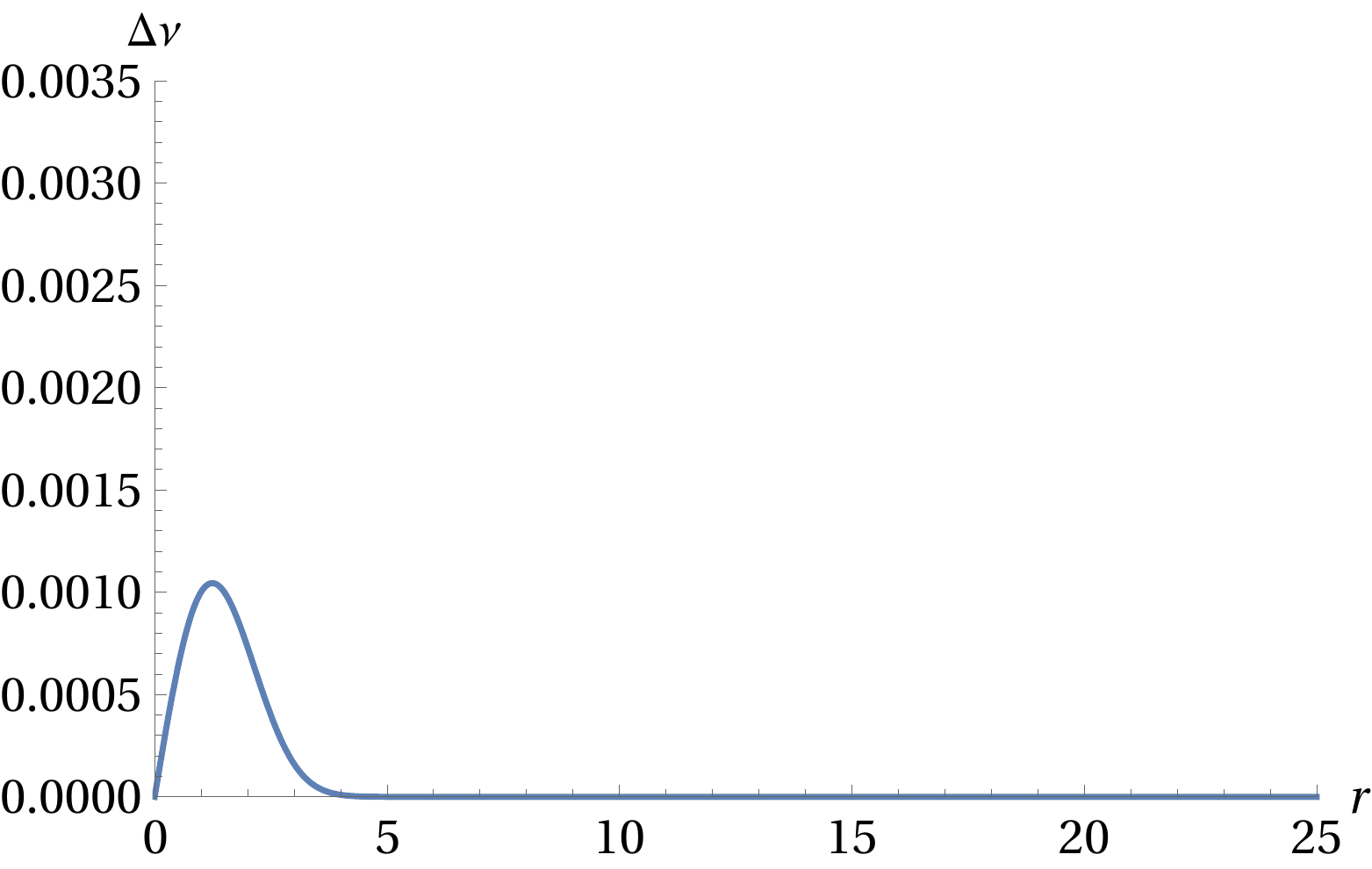}

\caption{Plot (a) illustrates tangential velocity $v$ of circular orbits for the Gaussian matter distribution \eqref{WAC-Gaussian}. Two cases are compared: $q=0$ [in blue] and $q=-10$ [in orange], and our parameter choices are $\rho_0=10^{-6}$ and $\sigma = 1$ (with $G=c=1$). It should be mentioned that for $|q| \neq 0$, $v(r)$ generally becomes imaginary for values of $r>0$ less than some value. Plot (b) illustrates the difference in rotation curves, where $\Delta v = v_{\text{GR}} -  v_{\text{MEMe}}$.}
\label{Fig:CurvesGaussian}
\end{figure}

\noindent The MEMe model is a conservative theory in the sense of \cite{Will2018}, as the standard PPN parameters are the same as that of GR. The extra parameters in the extended PPN formalism have the values given in Eqs. (\ref{GCA-Coefficients49a})--(\ref{GCA-NewPPNparamMEMe}), which differ from that of GR, so one expects circular orbits in the MEMe model to differ from those of GR, given some profile for the matter distribution. We first consider a Gaussian profile:
\begin{equation}\label{WAC-Gaussian}
\rho^* = \rho_0 \, e^{-{r^2}/{2 \sigma^2}},
\end{equation}

\noindent with $\rho_0$ being the central density and $\sigma$ a characteristic scale. Equation \eqref{WAC-PoissonEquationSphericalSolution} may be used to obtain the potentials:
\begin{align}
U =& \frac{2 \sqrt{2} \pi^{3/2} \, G \, \rho_0 \, \sigma^3}{r} \, \text{erf}\left[\frac{r}{\sqrt{2} \sigma }\right] , \nonumber \\
\Phi_2 =& -\frac{4 \pi^{5/2} \, G^2 \, \rho_0^2 \, \sigma^4}{r} \biggl\{ \text{erf}\left[\frac{r}{\sqrt{2} \sigma }\right] \biggl(\sqrt{\pi } \, r \, \text{erf} \left[\frac{r}{\sqrt{2} \sigma }\right] \nonumber \\
& \qquad \qquad \qquad \qquad + 2 \sqrt{2} \, \sigma \, e^{-\frac{r^2}{2 \sigma ^2}}\biggr) - 2 \sigma \, \text{erf}\left[\frac{r}{\sigma }\right]\biggr\} , \nonumber \\
\Psi =& \frac{\pi^{3/2} \, G^2 \, \rho_0^2 \, \sigma^3}{r} \, \text{erf}\left[\frac{r}{\sigma }\right] ,
\end{align}

\noindent which may be used to compute the tangential velocity $v(r)$ as given by Eq. \eqref{WAC-SpecificEnergyCircularOrbit3}. It turns out that a large modulus for $q$ is required to obtain rotation curves that differ from $q=0$ in a discernible way. For the Gaussian model, the tangential velocity of a circular orbit as a function of radius (rotation curve) is plotted in Fig. \ref{Fig:CurvesGaussian}, for the parameter choices $\rho_0=10^{-6}$ and $\sigma = 1$ (with $G=c=1$),
with one curve corresponding to $q=0$ and another corresponding to $q=10$. The rotation curve for $q=10$ is virtually identical to that of $q=0$ at large radii (as illustrated in the plot for the difference $\Delta v := v_{\text{GR}} -  v_{\text{MEMe}}$) and has an increased value for relatively small values of $r$. One might expect this behavior; for instance, one may note that $c_0 \Delta U \propto - q \rho > 0$ (for $q<0$) and upon comparison, one finds that the slope for $c_0 \Delta U (r) \propto \rho^*(r)$ [as given by Eq. \eqref{WAC-Gaussian}] matches the slope for the potential $U(r)$;
it follows that the counterterms enhance the force in the radial direction, which in turn increases $v(r)$. The convergence to the GR rotation curve at large $r$ is expected, as one expects the MEMe model to converge to GR at low density. These general features persist in the other examples we consider.

Another relevant matter profile is the isothermal one:
\begin{equation}\label{WAC-Isothermal}
\rho^* = \frac{M_h}{4 \, \pi \, a_h \, r^2}.
\end{equation}

\begin{figure}[htbp]

\vspace{-0.25cm}

\includegraphics[width=0.48 \textwidth]{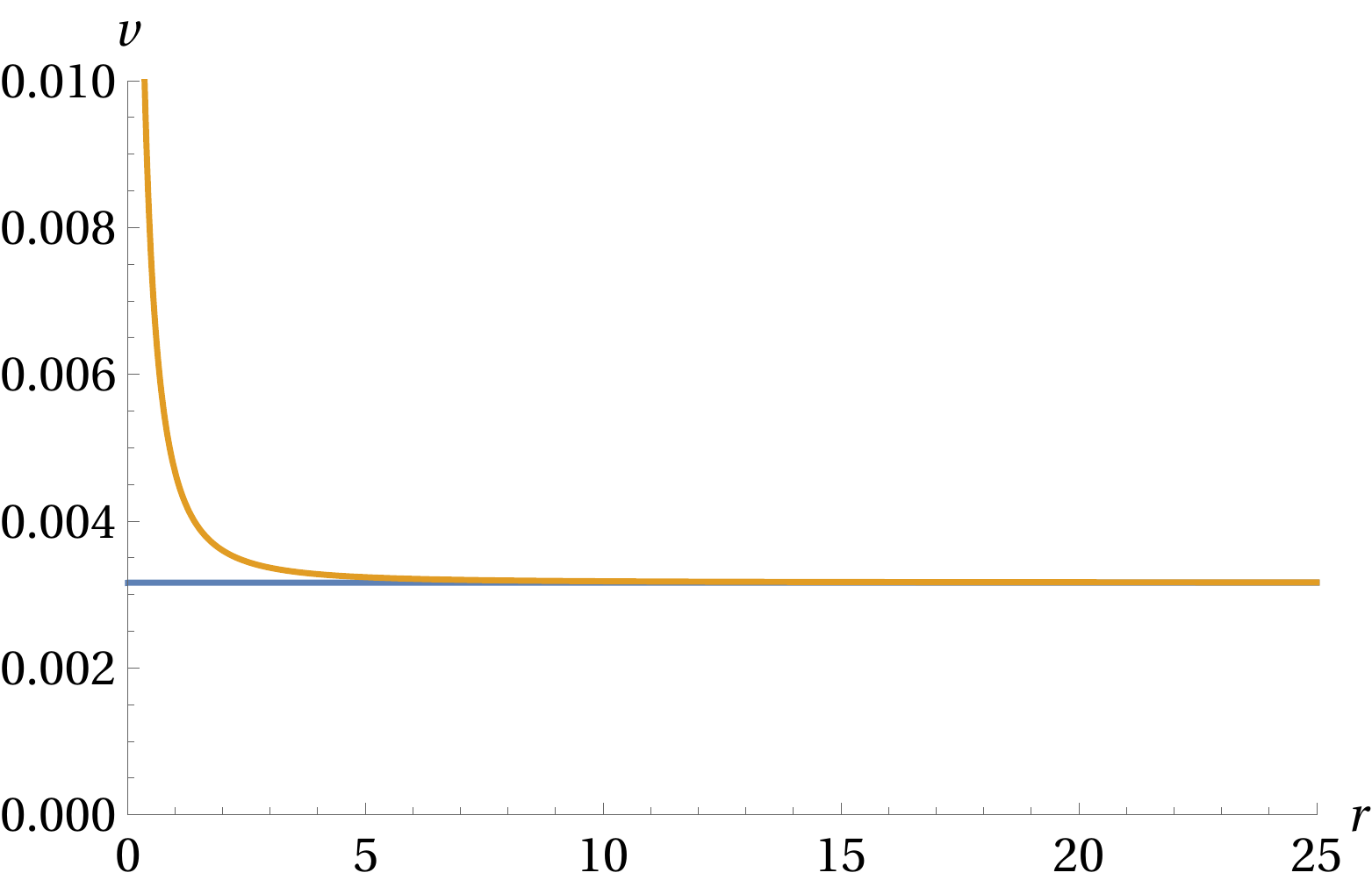}

\caption{This plot illustrates tangential velocity $v$ of circular orbits for the isothermal matter distribution \eqref{WAC-Isothermal}. Two cases are compared: $q=0$ [in blue] and $q=-10$ [in orange], and the parameter choices here are $M_h=10^{-2}$ and $a_h=10^{3}$ (with $G=c=1$). Again, as in Fig. \ref{Fig:CurvesGaussian}, for $q \neq 0$, $v(r)$ becomes imaginary for values of $r>0$ less than some value.}
\label{Fig:CurvesIsothermal}
\end{figure}

\begin{figure}[htbp]

\includegraphics[width=0.48 \textwidth]{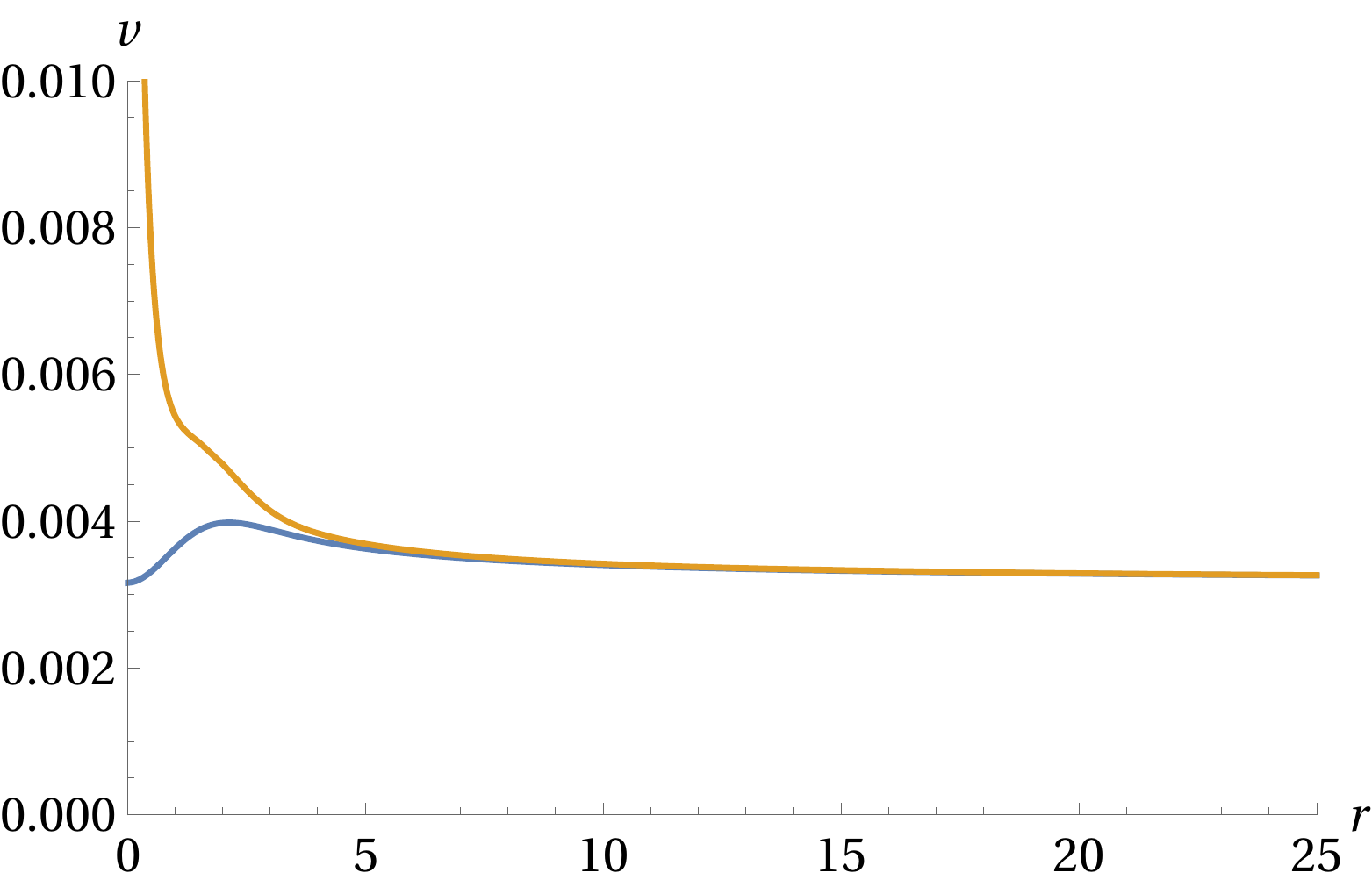}

\caption{Rotation curves for the combined Gaussian \eqref{WAC-Gaussian} and isothermal matter distributions \eqref{WAC-Isothermal}, using the same parameter choices as in Figs. \ref{Fig:CurvesGaussian} and \ref{Fig:CurvesIsothermal}.}
\label{Fig:CurvesGaussianIso}
\end{figure}

\noindent Such a profile is known to yield flat rotation curves in Newtonian gravity and is of interest (upon regularization of the singularity at $r=0$) for modeling dark matter halos. The curve $v(r)$ is plotted in Fig. \ref{Fig:CurvesIsothermal} for the parameter choices $M_h=10^{-2}$ and $a_h=10^{3}$. Again, one sees behavior similar to that of the Gaussian case---the $q = -10$ curve only differs (and has a lower value) from the $q=0$ case at small values for $r$, as expected. The divergence in the rotation curves at small $r$ is expected, since $\rho^*$ diverges in the limit $r \rightarrow 0$. In Fig. \ref{Fig:CurvesGaussianIso}, we plot $v(r)$ for the combined Gaussian and isothermal matter distributions
\begin{equation}\label{WAC-Combined}
\rho^* = \rho_0 \, e^{-{r^2}/{2 \sigma^2}} + \frac{M_h}{4 \, \pi \, a_h \, r^2},
\end{equation}

\noindent with the same parameter values as those of Figs. \ref{Fig:CurvesGaussian} and \ref{Fig:CurvesIsothermal}. Again, we note the velocities are increased at small $r$.

In all cases, we find that while the Laplacian counterterms have some effect on the behavior of rotation curves, the value of $q$ must be rather large in order to distinguish the MEMe model and GR, and even then, this occurs only at small values of $r$, as illustrated in the plots for $\Delta v$. If one expects the MEMe model to break down at the TeV scale, then $1/|q|$ is expected to be 30 orders of magnitude larger than the average density of Earth; for realistic astrophysical systems (galaxies), one might expect $1/|q|$ and the matter density to differ by an even greater amount. For the Gaussian example, the central density $\rho_0$ in Fig. \ref{Fig:CurvesGaussian} is $6$ orders of magnitude below the density scale $1/|q| = 10^{-1}$ at which the MEMe model breaks down. For the isothermal example, the average density $3M_h/4\pi a_h^3$ is $11$ orders of magnitude below the density scale $1/|q| = 10^{-1}$.

While these results suggest that signatures of the MEMe model are unlikely to appear in galactic rotation curves and dilute matter distributions, the MEMe model may still produce measurable differences in the interiors of neutron stars. The density for a neutron star is roughly an order of magnitude less than the highest energy-density ($\sim 14 \text{GeV}/\text{fm}^3 \approx 2.2 \times 10^{36} \text{J}/\text{m}^3$) states of matter probed to date in accelerator experiments \cite{Pasechnik2017,Mitchell2016}. If the scale for the cutoff density is assumed to be an order of magnitude higher than that of the quark-gluon density, so that it is $2$ orders of magnitude higher than the neutron star density, then upon modeling a neutron star with a Gaussian matter distribution, the term $c_0 \Delta U$ can become comparable to $-2 \Phi_2$ deep within the distribution. In particular, one can choose $\rho_0=M/2\sqrt{2} \pi^{3/2} \sigma^3$, with the normalization $M=G=1$ and $\sigma = 6$. In this case, the magnitude of the counterterm $c_0 \Delta U$ is roughly $\sim 0.75$ of the post-Newtonian correction $-2 \Phi_2$ when $r = \sigma/10$, though at the same radius, one finds $-c_0 \Delta U/2U^2 \sim 1.2 \times 10^{-3}$ and  $-c_0 \Delta U/2U \sim 1.7 \times 10^{-4}$, so the corrections are still rather small. However, this rough calculation suggests that the corrections from the MEMe model may modify the properties of the Neutron star in a measurable way.


%
%

%
%
\section{Summary and discussion}
\label{sec:summary}

In this article, we have extended the PPN formalism to handle a subclass of type-I MMGs and GCTs, and have applied the extended formalism to the MEMe model. Outside matter sources the Einstein frame and the Jordan frame coincide with each other and the field equations in either frame agree with those in GR. However, in the nonvacuum case, a PPN analysis for GCTs and the MEMe model should be performed with respect to the Jordan-frame metric $\mathfrak{g}_{\mu \nu}$. In fact, matter is minimally coupled to the Jordan-frame metric $\mathfrak{g}_{\mu \nu}$, and it is in this sense that the Jordan-frame metric is the physical metric. In order to perform a PPN analysis for $\mathfrak{g}_{\mu \nu}$, it is necessary to introduce an additional (dimensionful) potential $\Psi$ and counterterms (the latter vanish outside a matter distribution) constructed from the Laplacians of the PPN potentials. This can be understood considering the form of the field equations in the Jordan frame, which contains the energy density and its derivative up to the second order.  We have found that with the exception of the counterterm parameters and the parameter $\nu$ associated with $\Psi$, the parameters in the extended PPN formalism are the same as those of GR.

The new potential $\Psi$ and its associated parameter $\nu$ are not dimensionless. One might ask whether it is possible to define a dimensionless potential from $\Psi$. This can be done by choosing an appropriate length scale; however, such a procedure is not necessarily model independent. For example, to post-Newtonian order, a theory having the form of Eq. \eqref{GCA-AGeneralCouplingModel} would necessarily include a ${\rho^*}^2$ term on the rhs, the coefficient of which would introduce an additional scale. Indeed, each of the additional coefficients appearing on the rhs will introduce additional scales, and any of these can provide a reference scale to make $\Psi$ dimensionless. To avoid the choice of one scale rather than the other, here we have chosen to leave $\Psi$ and $\nu$ dimensionful.

Given some compact, spherical matter distribution, we have considered the monopole term in a standard multipole expansion and have found that to post-Newtonian order, the MEMe model is indistinguishable from GR in vacuum regions outside the matter distribution. This is not particularly surprising, as the Einstein- and Jordan-frame metrics coincide in vacuum, and one can for a single fluid in the Einstein frame absorb the differences from GR by a redefinition of fluid density and pressure. However, the differences between MEMe and GR become apparent when the details of the matter distribution are taken into account. The monopole expansion indicates that in MEMe, the effective gravitational constant $G$ depends on the internal structure of the source masses, and we argue that one can use this dependence to place strong constraints on the free parameter $q$ of MEMe. In particular, we argue that (conceptual issues aside; see the next paragraph) a modification of the experiment described in \cite{Lietal2018} may improve the constraint on $q$ over the speed of light constraint of \cite{FengCarloni2019} by 10 orders of magnitude. In particular, we propose an experiment in which the spherical source masses are disassembled into concentric ``thick'' shells, and the active gravitational masses of the individual shells and the assembled spheres are compared.

This proposal might bring up a conceptual issue regarding the gravitational binding energy between concentric thick shells of matter. In GR, this situation can be treated using the standard junction and thin-shell formalism of Israel \cite{Israel1966,*Poisson}. Since the geometry outside the shells is essentially that of GR, one might ask whether the binding energy is modified at all. This question depends on the behavior of the theory at the boundaries of spatially compact matter distributions, which can be rather subtle in certain theories of modified gravity. In the case of EiBI gravity \cite{BanadosFerreira2010}, which shares a structure similar to that of the MEMe model in the weak-field limit [it falls into the class of models described by Eq. \eqref{GCA-AGeneralCouplingModel} and has a Newtonian potential resembling Eq. \eqref{NLJF}], it was argued in \cite{PaniSotiriouPRL2012} that discontinuities in matter distributions, such as those at the boundaries of stars, can generate unacceptable curvature singularities in EiBI gravity. However, we have argued that in the Newtonian limit of the MEMe model, such singularities correspond to strong gravitational forces acting on matter which lead to a rearrangement of matter distributions, so that the gravitational backreaction may resolve such singularities---similar arguments have been made for EiBI theory \cite{Kim2014} (see also \cite{BeltranJimenezetal2017}). A detailed investigation of this issue beyond the Newtonian limit in the MEMe model will be left for future work.

Finally, we compared the post-Newtonian predictions of the MEMe model and GR within a matter distribution to understand the effects of the counterterms that appear in the gravitational potential. In particular, we studied the behavior of circular geodesics in the presence of spherically symmetric Gaussian and isothermal matter distributions. Plots of the tangential velocity rotation curves indicate that the predictions of the MEMe model only differ significantly from that of GR only for high matter densities and large values for the parameter $q$. It follows from this result that the MEMe model alone cannot describe galactic rotation curves in the absence of dark matter---in fact, the MEMe model (slightly) \textit{increases} orbital velocities at small radii---and the differences in the behavior of geodesics between the MEMe model and GR are minimal even within a distribution of dark matter. These results also indicate that, in general, the counterterms do not have a strong effect on the geodesics unless the parameter $q$ is increased to an unrealistically large value. On the other hand, a rough estimate suggests that, for a cutoff density $1/q$ an order of magnitude higher than the highest densities probed in accelerator experiments, the MEMe model may yield measurable corrections to the properties of neutron stars.

In the present paper, we have considered the MEMe model as a type-I MMG theory and have focused on its gravitational aspects. Alternatively, in the Einstein frame, one can consider the MEMe model as a theory of a modified matter action minimally coupled to GR. Indeed, after integrating out the auxiliary tensor field $A{_\mu}{^\alpha}$ the matter action in the Einstein frame is modified in such a way that the fields in the standard model of particle physics acquire additional (renormalizable and nonrenormalizable) interactions among themselves. In future work, it is certainly interesting to study phenomenological consequences of those extra interactions (that remain even in the $G\to 0$ limit) and their implications to collider physics, cosmic rays, early Universe cosmology, and so on.

The extended PPN formalism developed in the present paper may be applied to some of other type-I MMG theories. It is worthwhile investigating the PPN constraints on theories in this class and also extending the formalism so that it can be applied to other type-I MMG theories and some type-II MMG theories as well.


%
%


\begin{acknowledgments}
We would like to thank Vitor Cardoso and Chinmoy Bhattacharjee for helpful comments. Some of the calculations were performed using the xAct package \cite{xActbib} for \textit{Mathematica}. J.C.F. acknowledges support from Funda\c{c}\~{a}o para a Ci\^{e}ncia e a Tecnologia Grants No. PTDC/MAT-APL/30043/2017 and No. UIDB/00099/2020. The work of S.M. was supported in part by Japan Society for the Promotion of Science Grants-in-Aid for Scientific Research No.~17H02890, and No.~17H06359 and by World Premier International Research Center Initiative, The Ministry of Education, Culture, Sports, Science and
Technology, Japan.
\end{acknowledgments}


\bibliography{PPNMEMe}

\end{document}